\documentclass[journal]{IEEEtran}
\usepackage{amsmath,amsfonts,amssymb}
\usepackage{algorithmic}
\usepackage{algorithm}
\usepackage{array}
\usepackage[caption=false,font=footnotesize,labelfont=rm,textfont=rm]{subfig}
\usepackage{subfig}
\usepackage{textcomp}
\usepackage{stfloats}
\usepackage{url}
\usepackage{verbatim}
\usepackage{graphicx}
\usepackage{cite}
\usepackage{balance}
\usepackage{color, xcolor}

\begin{document}

	\title{Movable Antenna-Enabled Co-Frequency Co-Time Full-Duplex Wireless Communication}

	\author{
		Jingze Ding, \IEEEmembership{Graduate Student Member, IEEE},
		Zijian Zhou, \IEEEmembership{Member, IEEE},
		Wenyao Li,
		Chenbo Wang, \IEEEmembership{Graduate Student Member, IEEE},
		Lifeng Lin, and
		Bingli Jiao, \IEEEmembership{Senior Member, IEEE}

		\thanks{This work was supported in part by Science and Technology Project of Guangzhou under Grants 202206010118 and 2023B04J0011 and in part by National Natural Science Foundation of China under Grant 62171006. The calculations were supported by High-Performance Computing Platform of Peking University. The associate editor coordinating the review of this letter and approving it for publication was Carmen D'Andrea. \emph{(Corresponding author: Zijian Zhou.)}}
		\thanks{The authors are with School of Electronics, Peking University, Beijing 100871, China (e-mail: djz@stu.pku.edu.cn; zjzhou1008@pku.edu.cn; liwenyao@stu.pku.edu.cn; \{wcb15, linlifeng, jiaobl\}@pku.edu.cn).}}

	\maketitle

	\begin{abstract}
		Movable antenna (MA) provides an innovative way to arrange antennas that can contribute to improved signal quality and more effective interference management.  This technology is especially beneficial for co-frequency co-time full-duplex (CCFD) wireless communication, which struggles with self-interference (SI) that usually overpowers the desired incoming signals.  By dynamically repositioning transmit/receive antennas, we can mitigate the SI and enhance the reception of incoming signals.  Thus, this paper proposes a novel MA-enabled point-to-point CCFD system and formulates the minimum achievable rate of two CCFD terminals.  To maximize the minimum achievable rate and determine the positions of MAs, we introduce a solution based on projected particle swarm optimization (PPSO), which can circumvent common suboptimal positioning issues.  Moreover, simulation results reveal that the PPSO method leads to better performance compared to the conventional alternating position optimization (APO).  The results also demonstrate that an MA-enabled CCFD system outperforms the one using fixed-position antennas (FPAs).
	\end{abstract}


	\section{Introduction}
	\IEEEPARstart{W}{ireless} communications are experiencing a dramatic increase in the demand for never-enough capacity.  To meet this demand, numerous empowering technologies such as millimeter-wave (mmWave) and co-frequency co-time full-duplex (CCFD) have been proposed \cite{mmWave_2024, Liu_Survey_2015, Guan_WCL_2022}.  Among these, CCFD stands out for its potential to double spectral efficiency by enabling simultaneous information exchange over the same frequency band.  However, the implementation of this technology encounters a significant challenge due to the presence of self-interference (SI) leaking from the transmit antenna.

	In the domain of antenna, conventional methods mitigate the SI using passive isolation techniques such as separate distance, orientation, and polarization \cite{Duarte_TWC_2012,Dash_AWPL_2022}.  These often rely on fixed-position antennas (FPAs), which limit the extensive exploitation of spatial degrees of freedom (DoFs) in adapting to channel variations.  To overcome this limitation, movable antenna (MA) \cite{Zhu_TWC_2023, Zhu_CommMag_2023} has recently emerged as a promising solution.  It allows flexible adjustments of antennas' positions within a specified region and enables the improvement or degradation of the channel condition.  Each MA is connected to the radio frequency chain via a flexible cable and mounted on a mechanical slide, which can be controlled by step motors for free movement.  Alternatively, micro-electromechanical systems (MEMS) can offer more precise control, especially for smaller antennas \cite{Zhu_TWC_2023, Zhu_CommMag_2023, Book_MEMS_2023}.  While the moving region is sufficient, MAs can outperform antenna selection (AS) \cite {AS_method}, another method that exploits the spatial DoFs.

	Several studies have initially explored MA-enabled wireless communication systems, e.g., multiple-input multiple-output (MIMO) \cite{Ma_TWC_2023}, multiuser \cite{Xiao_arXiv_2023, Zhu_TWC_2023_1}, and secure communication systems \cite{MA_Secure1, MA_Secure2, MA_Secure2_Ding,MA_Secure2_Ding1, MA_Secure3}.  Nevertheless, in contrast to CCFD, most terminals in the aforementioned systems aided by MAs operate in half-duplex (HD) mode, which may not meet the demands for progressively scarce spectral resources.  The feature of MAs to leverage spatial DoFs is crucial in reducing interference and enhancing the signal of interest (SoI), making it well-suited for CCFD and multiuser systems. Unlike multiuser systems employing beamforming to suppress multiuser interference, CCFD systems need to suppress SI with much higher power than the SoI by SI cancellation modules. Hence, it is paramount that the analog and digital SI cancellations adapt to channel variations induced by the movement of MA.  From the knowledge of the authors, the issue of CCFD systems with MAs has been disregarded.

	Motivated by these considerations, we propose a novel MA-enabled CCFD system.  Our main contributions are threefold as follows.  First, we formulate the optimization problem by maximizing the minimum achievable rate of two CCFD terminals with the aid of MAs.  Next, to circumvent an undesired suboptimal positioning, we introduce the projected particle swarm optimization (PPSO) as a superior alternative to the conventional alternating position optimization (APO) method.  Last, we provide numerical results that not only illustrate a better performance of the PPSO method over traditional solutions but also substantiate the enhancement of the proposed MA-enabled CCFD system compared to systems with FPAs.

	\textit{Notation}: $a$, $\mathbf{a}$, and $\mathbf{A}$ denote a scalar, a vector, and a matrix, respectively.  ${\left(  \cdot  \right)^{\rm{T}}}$, ${\left(  \cdot  \right)^{\rm{H}}}$, and $\left| \cdot \right|$  stand for transpose, conjugate transpose, and absolute value, respectively. $\odot$ represents Hadamard product.  ${\left[ \mathbf{a} \right]_i}$ denotes the $i^\mathrm{th}$ element of the vector $\mathbf{a}$. $\mathbb{C}^{M \times N}$ and $\mathbb{R}^{M \times N}$ are the sets for complex and real matrices of $M \times N$ dimensions, respectively. The circularly symmetric complex Gaussian (CSCG) distribution with mean zero and variance $\sigma^2$ is represented by $\mathcal{CN}\left(0,\sigma^2\right)$.

	\section{System Model}
	\begin{figure}
		\centering
		\includegraphics[width=0.85\linewidth]{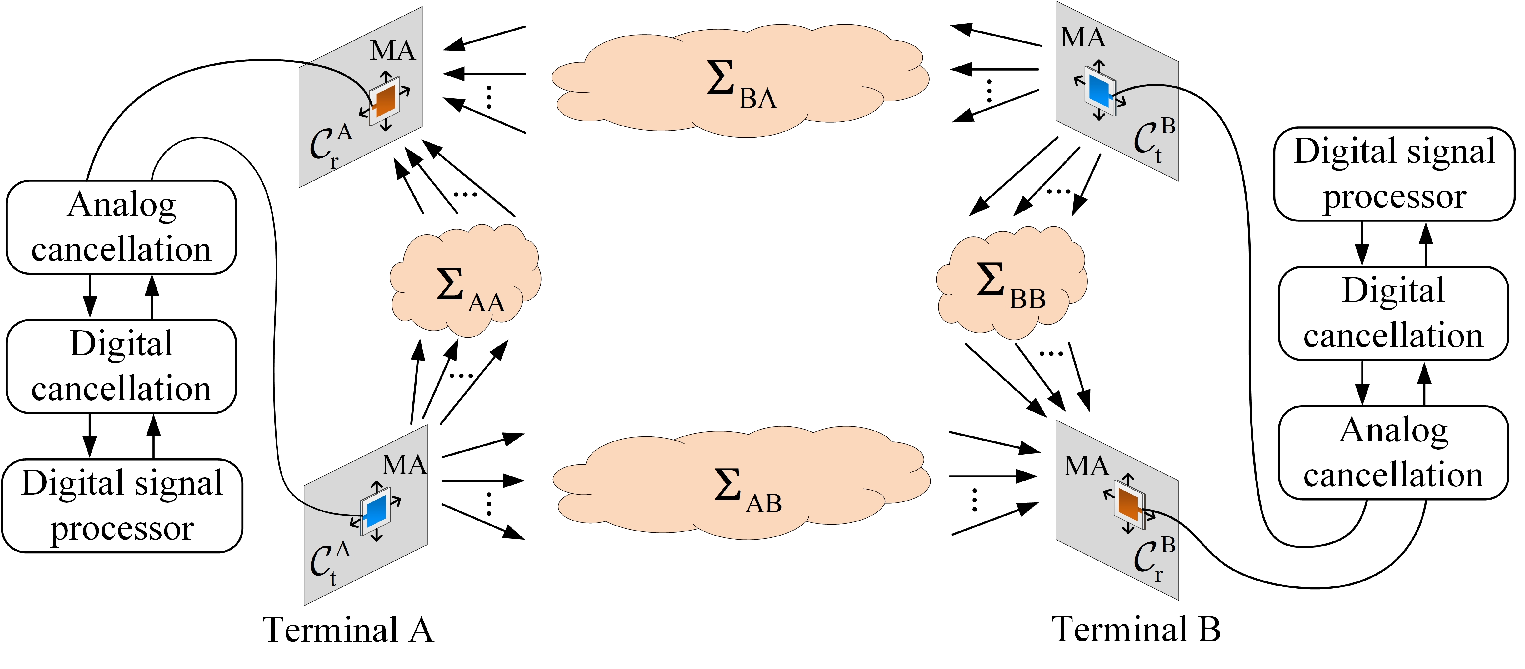}
		\caption{An MA-enabled point-to-point CCFD system.}
		\label{fig1}
	\end{figure}
	Fig.\;\ref{fig1} depicts an MA-enabled point-to-point CCFD system that comprises terminals A and B.  Each terminal is equipped with a single transmit MA and a single receive MA, which can be driven within a two-dimensional region.  The positions of transmit and receive MAs are denoted using Cartesian coordinates, i.e.,  $\mathbf{t}_p=\left[x_{\mathrm{t}}^p,y_{\mathrm{t}}^p\right]^{\mathrm{T}} \in \mathcal{C}_\mathrm{t}^p$ and $\mathbf{r}_q=\left[x_{\mathrm{r}}^q,y_{\mathrm{r}}^q\right]^{\mathrm{T}} \in \mathcal{C}_\mathrm{r}^q$, where $\mathcal{C}_\mathrm{t}^p$ and $\mathcal{C}_\mathrm{r}^q$ are the corresponding transmit and receive regions at terminals $p$ and $q$.  Herein, $p$ and $q$ belong to the set $\{\mathrm{A}, \mathrm{B}\}$, signifying the terminal labels.  This notation will be consistently used in the following derivations.

	In this MA-enabled CCFD system, the channel responses rely on the variable positions of MAs.  We thus characterize the channel coefficient as a function of the transmit and receive MAs' coordinates, i.e., ${h_{pq}}\left({\mathbf{t}_p},{\mathbf{r}_q}\right)$\footnote{It is worth noting that our proposed system can be expanded to the MIMO system. The channel matrix of the MIMO system with $M$ transmit and $N$ receive MAs can be given by $\mathbf{H}\left( {\tilde {\mathbf{t}}}, {\tilde {\mathbf{r}}} \right) \in\mathbb{C}^{N \times M}$, where ${\tilde {\mathbf{t}}} = \left[\mathbf{t}_1, \cdots, \mathbf{t}_M\right]\in\mathbb{R}^{2 \times M}$ and ${\tilde {\mathbf{r}}} = \left[\mathbf{r}_1, \cdots, \mathbf{r}_N\right]\in\mathbb{R}^{2 \times N}$.  Here, $\mathbf{t}_m$ ($1 \le m \le M$) and $\mathbf{r}_n$ ($1 \le n \le N$) represent the coordinates of $m^\mathrm{th}$ transmit and $n^\mathrm{th}$ receive MAs, respectively.}.  Suppose that the channel state information (CSI) is perfectly known at both terminals\footnote{Although acquiring the perfect CSI is challenging, existing works \cite{Ma_CL_2023,Xiao_arXiv_2023_1} have proposed some practical methods with low pilot overhead and computational complexity to achieve satisfactory CSI estimation. Besides, our proposed scheme with full CSI can provide a performance upper bound for realistic scenarios and robust designs.}.  Denoting the transmit power of the terminals as $P_\mathrm{t}$, the received signals at terminal $q$ can be expressed as
	\begin{align} \label{2-1}
		z_q \left(\mathbf{t}_p, \mathbf{r}_q\right) & = h_{pq} \left(\mathbf{t}_p, \mathbf{r}_q\right) \sqrt{P_\mathrm{t}} s_p  \nonumber \\
		&+ h_{qq} \left(\mathbf{t}_q, \mathbf{r}_q\right) \sqrt{P_\mathrm{t}} s_q + n_q,~p\ne q.
	\end{align}
	Here, $s_p$ and $s_q$ stand for the transmitted signals with zero mean and normalized power of one.  $n_q\sim{\mathcal{CN}}\left(0,\sigma^2_q\right)$ is the additive white Gaussian noise (AWGN) with the power of $\sigma^2_q$.  Besides, the first term in \eqref{2-1} represents the SoI, while the second term is the inherent SI in CCFD systems.

	\begin{figure}
		\centering
		\includegraphics[width=0.8\linewidth]{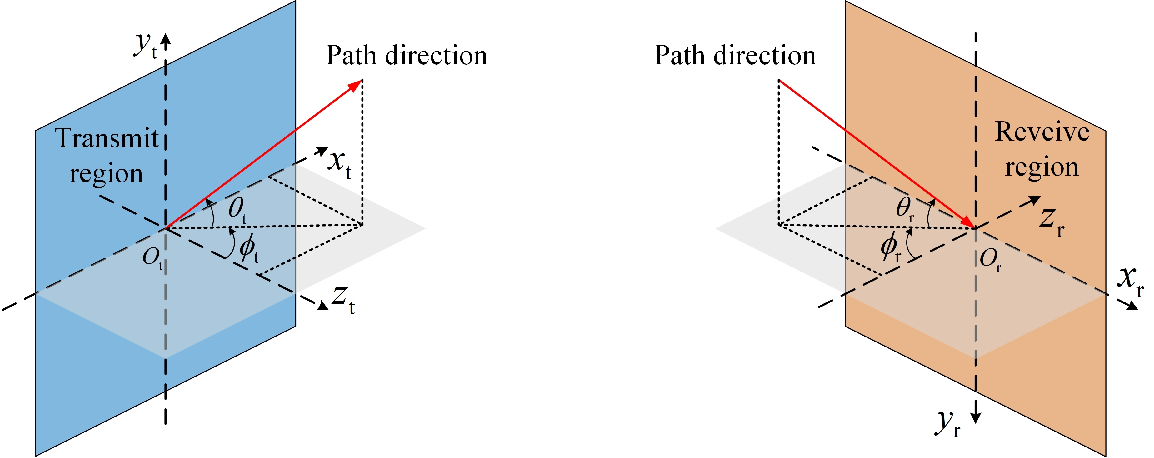}
		\caption{The coordinates and spatial angles for transmit and receive MAs.}
		\label{fig2}
	\end{figure}

	\subsection{Channel Model}
	According to the field-response channel model in \cite {Zhu_TWC_2023}, we consider quasi-static block-fading channels, focusing on a specific fading block where the multipath channel components at any location in the transmit/receive region are fixed \cite{MA_Secure2}.  Denote $L^{pq}_\mathrm{t}$ and $L^{pq}_\mathrm{r}$ as the number of transmit and receive paths, respectively. Compared to the origin coordinate, the difference of propagation distance for transmit MA at the $l_\mathrm{t}^\mathrm{th}$ transmit path ($1 \le l_\mathrm{t} \le L^{pq}_\mathrm{t}$) can be expressed as $\rho _{\mathrm{t},l_\mathrm{t}}^p\left(\mathbf{t}_p\right) = x_\mathrm{t}^p\cos \theta _{\mathrm{t},l_\mathrm{t}}^p\sin \phi _{\mathrm{t},l_\mathrm{t}}^p + y_\mathrm{t}^p\sin \theta _{\mathrm{t},l_\mathrm{t}}^p$, where $\theta _{\mathrm{t},l_\mathrm{t}}^p \in \left[ { - \frac{\pi}{2},\frac{\pi}{2}} \right]$ and $\phi _{\mathrm{t},l_\mathrm{t}}^p \in \left[ { - \frac{\pi}{2},\frac{\pi}{2}} \right]$ respectively denote the elevation and azimuth angles of departure (AoDs), as shown in Fig.\;\ref{fig2}. Similarly, denote the elevation and azimuth angles of arrival (AoAs) at the $l_\mathrm{r}^\mathrm{th}$ receive path ($1 \le l_\mathrm{r} \le L^{pq}_\mathrm{r}$) as $\theta _{\mathrm{r},l_\mathrm{r}}^q \in \left[ { - \frac{\pi}{2},\frac{\pi}{2}} \right]$ and $\phi _{\mathrm{r},l_\mathrm{r}}^q \in \left[ { - \frac{\pi}{2},\frac{\pi}{2}} \right]$, respectively. The difference of propagation distance for receive MA at the $l_\mathrm{r}^\mathrm{th}$ receive path is given by $\rho _{\mathrm{r},l_\mathrm{r}}^q\left(\mathbf{r}_q\right) = x_\mathrm{r}^q\cos \theta _{\mathrm{r},l_\mathrm{r}}^q\sin \phi _{\mathrm{r},l_\mathrm{r}}^q + y_\mathrm{r}^q\sin \theta _{\mathrm{r},l_\mathrm{r}}^q$.

	Suppose that the far-field condition is satisfied and the changes in AoDs, AoAs, and amplitudes for the multiple channel paths are omitted for any given $\mathbf{t}_p$ and $\mathbf{r}_q$.  Thus, we only consider the phase difference \cite {Zhu_TWC_2023}, which is written by $\frac{2\pi \rho}{\lambda}$.  Here, $\rho$ is the difference of propagation distance and $\lambda$ is the wavelength.  The field-response vectors for positions $\mathbf{t}_p$ and $\mathbf{r}_q$ are respectively given by
	\begin{equation} \label{2-4}
		\mathbf{g}\left(\mathbf{t}_p\right) = \left[ e^{j\frac{2\pi}{\lambda} \rho_{\mathrm{t},1}^p \left(\mathbf{t}_p\right)}, \cdots, e^{j\frac{2\pi}{\lambda} \rho_{\mathrm{t},L_\mathrm{t}^{pq}}^p \left(\mathbf{t}_p\right)} \right]^\mathrm{T} \in {\mathbb{C}^{{{L}_\mathrm{t}^{pq}} \times 1}},
	\end{equation}
	and
	\begin{equation} \label{2-5}
		\mathbf{f}\left(\mathbf{r}_q\right)  = \left[ e^{j\frac{2\pi}{\lambda} \rho_{\mathrm{r},1}^q \left(\mathbf{r}_q\right)}, \cdots, e^{j\frac{2\pi}{\lambda} \rho_{\mathrm{r},L_\mathrm{r}^{pq}}^q \left(\mathbf{r}_q\right)} \right]^\mathrm{T} \in {\mathbb{C}^{{{L}_\mathrm{r}^{pq}} \times 1}}.
	\end{equation}

	As a result, the channel coefficient between the transmit and receive MAs can be obtained by
	\begin{equation} \label{2-6}
		h_{pq}\left({\mathbf{t}_p},{\mathbf{r}_q}\right) = \mathbf{f}\left( {\mathbf{r}_q}\right)^{\mathrm{H}}\mathbf{\Sigma}_{pq} \mathbf{g}\left( {\mathbf{t}_p}\right),
	\end{equation}
	where $\mathbf{\Sigma}_{pq} \in {\mathbb{C}^{{{L}_\mathrm{r}^{pq}} \times{{L}_\mathrm{t}^{pq}}}}$ is the channel response from the origin of the transmit region to that of the receive region\footnote{Note that when $p = q$ and $p \ne q$, the channel coefficient $h_{pq}$ can be interpreted as the SI channel and the SoI channel, respectively.}.

	\subsection{Problem Formulation}
	For a CCFD system, the signal-to-interference-plus-noise ratio (SINR) is an important metric to evaluate the system's performance.  In this scenario, it can be represented by
	\begin{equation} \label{2-7}
		{\gamma^q} = \frac{{{{\left| {{h_{pq}}\left( {\mathbf{t}_p},{\mathbf{r}_q}\right) } \right|}^2}{P_\mathrm{t}}}}{{{{\left| {{h_{qq}}\left( {\mathbf{t}_q},{\mathbf{r}_q}\right) } \right|}^2}{P_\mathrm{t}} + \sigma_q^2}},~p\ne q.
	\end{equation}
	Therefore, based on \eqref{2-7}, the achievable rate at terminal $q$ is calculated by $R_q = {\log_2}\left( 1 + {\gamma^q}\right)$.

	To ensure fairness between the terminals, we aim to maximize the minimum achievable rate by jointly optimizing the positions of the four MAs.  To this end, the optimization problem is formulated as
	\begin{align}
		\mathop{\max} \limits_{\mathbf{t}_p, \mathbf{r}_q} \quad
		&\min\{{R_\mathrm{A}}, {R_\mathrm{B}}\} \label{2-9}\\
		\mathrm{s.t.} \quad
		&\mathbf{t}_p \in \mathcal{C}_\mathrm{t}^p,~ \mathbf{r}_q \in \mathcal{C}_\mathrm{r}^q, ~ p,q \in\left\{ \mathrm{A,B}\right\}. \tag{6a} \label{2-9a}
	\end{align}

	Constraint \eqref{2-9a} indicates that MAs are confined to movement within a determined transmit/receive region.  Even though problem \eqref{2-9} appears simple, it is difficult to solve due to the non-convex objective function.  Moreover, the position of each MA has a significant impact not only on the reception of the SoI but also on the SI, thereby leading to a strong interrelationship among the positions of the four MAs.

	\section{Proposed Solution}
	Since there are four highly coupled positions of MAs, the conventional APO optimizing one of them with the other three being fixed risks converging to an undesired suboptimal solution \cite{Xiao_arXiv_2023}. A intuitive approach to avoid it is to use a traversal algorithm, at the sacrifice of higher complexity. It is worth noting that the PPSO method \cite{Xiao_arXiv_2023, MA_Secure2_Ding} can simultaneously optimize the positions of all MAs to avoid undesired suboptimal solutions, while being much less complex than the traversal algorithm.  Considering these factors, we will focus on the PPSO-based solution in this section.

	\subsection{PPSO Method}
	To begin with, denoting $N$ as the number of particles and $K$ as the number of iterations, the positions of particles during the $k^\mathrm{th}$ iteration ($1 \le k \le K$) are represented by a position matrix, i.e., ${\mathbf{U}^{(k)}} = \left[ {\mathbf{u}_1^{(k)},\cdots,\mathbf{u}_N^{(k)}} \right] \in {\mathbb{R}^{8 \times N}}$. The position vector of the $n^\mathrm{th}$ particle ($1 \le n \le N$) is defined as
	\begin{equation} \label{3-1}
		\mathbf{u}_n^{(k)} = {\left[ {\mathbf{t}_{\mathrm{A},n}^{(k)}}^\mathrm{T}, {\mathbf{r}_{\mathrm{A},n}^{(k)}}^\mathrm{T}, {\mathbf{t}_{\mathrm{B},n}^{(k)}}^\mathrm{T}, {\mathbf{r}_{\mathrm{B},n}^{(k)}}^\mathrm{T} \right]^\mathrm{T}} ,
	\end{equation}
	where $\mathbf{t}_{p,n}^{(k)}$ and $\mathbf{r}_{q,n}^{(k)}$ represent the transmit and receive MAs' coordinates, respectively, and $\mathbf{u}_n^{(k)}$ might be a potential solution to problem \eqref {2-9}.  The velocity matrix of the particles is introduced as ${\mathbf{V}^{(k)}} = \left[ {\mathbf{v}_1^{(k)}, \cdots ,\mathbf{v}_N^{(k)}} \right] \in {\mathbb{R}^{8 \times N}}$.

	For constraint \eqref{2-9a}, the coordinates $\mathbf{t}_{p,n}^{(k)}$ and $\mathbf{r}_{q,n}^{(k)}$ in \eqref{3-1} are restricted to a given region.  We assume $\mathcal{C}_\mathrm{t}^p$ and $\mathcal{C}_\mathrm{r}^q$ are square regions, each with size $D\times D$.  Therefore, the vector $\mathbf{u}_n^{(k)}$ is updated as
	\begin{equation} \label{3-2}
		\mathbf{u}_n^{(k)} = \mathcal{D}\left\{ \mathbf{u}_n^{(k - 1)} + \mathbf{v}_n^{(k)}\right\},
	\end{equation}
	where $\mathcal{D}\left\lbrace \mathbf{a}\right\rbrace $ is a function that projects each component of vector $\mathbf{a}$ to its corresponding maximum/minimum value, i.e.,
	\begin{equation} \label {3-3}
		{\left[ {\mathcal{D}\left\{ \mathbf{a} \right\}} \right]_i} = \left\{
		\begin{array}{rcl}
			\frac{D}{2}, & \mathrm{if} & {{\left[ \mathbf{a} \right]_i}>\frac{D}{2}}\\
			{\left[ \mathbf{a} \right]_i}, & \mathrm{if} & {-\frac{D}{2} \leq {\left[ \mathbf{a} \right]_i} \leq \frac{D}{2}} \\
			-\frac{D}{2}, & \mathrm{if} & {{\left[ \mathbf{a} \right]_i} < -\frac{D}{2}} \\
		\end{array} \right..
	\end{equation}

	In the PPSO method, incorporating with \eqref{3-2}, the velocity vector of the $n^\mathrm{th}$ particle at the $k^\mathrm{th}$ iteration is calculated by
	\begin{align} \label{3-4}
		\mathbf{v}_n^{(k)} = \omega\mathbf{v}_n^{(k - 1)}  &+ {c_1}{\mathbf{e}_1}\odot \left(\mathbf{u}_n^\star - \mathbf{u}_n^{(k - 1)}\right) \nonumber \\
		&+ {c_2}{\mathbf{e}_2} \odot \left(\mathbf{u}^\star - \mathbf{u}_n^{(k - 1)}\right).
	\end{align}
	Here, the inertia weight, $\omega$, is pivotal in moderating the impact of the prior velocity, $\mathbf{v}_n^{(k - 1)}$, on the subsequent velocity, $\mathbf{v}_n^{(k)}$.  The learning factors, $c_1$ and $c_2$, serve as the step sizes that guide each particle towards the local optimal position vector, $\mathbf{u}_{n}^\star$, and the global optimal position vector, $\mathbf{u}^\star$.  Furthermore, the random vectors, $\mathbf{e}_1, \mathbf{e}_2 \in {\mathbb{R}^{8 \times 1}}$, where each entry is a uniform random number in the range $\left[0,1\right]$, are used to increase the randomness of the search for reducing the risk of obtaining an undesired local optimal solution. To balance the trade-off between the speed and precision of the PPSO search, $\omega$ is described as a linearly decreasing function across the iteration count in the interval $\left[ {{\omega _{\min }},{\omega _{\max }}} \right]$, calculated as $\omega  =  {\omega _{\max }} - \left( {{\omega _{\max }} - {\omega_{\min }}} \right) \frac{k}{K}$.

	\begin{algorithm}[!t]
		\caption{PPSO method}
		\label{alg1}
		\small
		\renewcommand{\algorithmicrequire}{\textbf{Input:}}
		\renewcommand{\algorithmicensure}{\textbf{Output:}}
		\begin{algorithmic}[1]
			\REQUIRE $\left\{ {{\mathcal{C}}_\mathrm{t}^p} \right\}$, $\left\{ {{\mathcal{C}}_\mathrm{r}^q} \right\}$, $\left\{ {\theta_{\mathrm{t},l_\mathrm{t}}^p} \right\}$, $\left\{ {\phi_{\mathrm{t},l_\mathrm{t}}^p} \right\}$, $\left\{ {\theta_{\mathrm{r},l_\mathrm{r}}^q} \right\}$, $\left\{ {\phi_{\mathrm{r},l_\mathrm{r}}^q} \right\}$, $N$, $K$, $c_1$, $c_2$, $\omega_{\min }$, $\omega_{\max}$.
			\ENSURE ${\mathbf{u}^\star}$.
			\STATE Randomly initialize the position matrix $\mathbf{U}^{(0)}$ and the velocity matrix $\mathbf{V}^{(0)}$.
			\STATE Initialize the local and global optimal position vectors, $\mathbf{u}^\star_n = \mathbf{u}_n^{(0)}$ and $\mathbf{u}^\star = \arg \mathop {\max }\limits_{\mathbf{u}_n^{(0)}} \left[ {F\left( {\mathbf{u}_1^{(0)}} \right), \cdots, F\left( {\mathbf{u}_N^{(0)}} \right)} \right]$, respectively.
			\FOR{$k = 1$ to $K$}
			\STATE Calculate the inertia weight $\omega$.
			\FOR{$n = 1$ to $N$}
			\STATE Calculate the position and velocity vectors of the $n^\mathrm{th}$ particle, $\mathbf{u}_n^{(k)}$ and $\mathbf{v}_n^{(k)}$, respectively.
			\STATE Update the fitness value of the $n^\mathrm{th}$ particle $F\left( {\mathbf{u}_n^{(k)}} \right)$.
			\IF{$F\left( {\mathbf{u}_n^{(k)}} \right) > F\left({\mathbf{u}_n^\star}\right)$}
			\STATE $\mathbf{u}_n^\star = \mathbf{u}_n^{(k)}$.
			\ENDIF
			\IF{$F\left( {\mathbf{u}_n^{(k)}} \right) > F\left({\mathbf{u}^\star}\right)$}
			\STATE $\mathbf{u}^\star = {\mathbf{u}_n^{(k)}}$.
			\ENDIF
			\ENDFOR
			\ENDFOR
			\RETURN $\mathbf{u}^\star$.
		\end{algorithmic}
	\end{algorithm}

	During each iteration, the fitness function determines the local and global optimal position vectors. Assuming that the optimal position vector corresponds to the largest fitness value.  To maximize the minimum achievable rate in \eqref{2-9}, the fitness function of each particle is established as
	\begin{equation} \label{3-6}
		F\left( {\mathbf{u}_n^{(k)}} \right) = \min \left\{ {{R_\mathrm{A}}\left( {\mathbf{u}_n^{(k)}} \right), {R_\mathrm{B}}\left( {\mathbf{u}_n^{(k)}} \right)} \right\},
	\end{equation}
	where ${R_\mathrm{A}}\left( {\mathbf{u}_n^{(k)}} \right)$ and ${R_\mathrm{B}}\left( {\mathbf{u}_n^{(k)}} \right)$ are the achievable rates of the terminals when the positions of MAs are specified by ${\mathbf{u}_n^{(k)}}$. Upon completing $K$ iterations, the near-optimal positions of the four MAs are obtained as ${\mathbf{u}^\star}$, in general. The PPSO method to address problem \eqref{2-9} is presented in Algorithm\;\ref{alg1}.

	\subsection{Convergence and Complexity Analysis}
	The convergence of the PPSO method is ensured for two key reasons. First, the fitness value of the global optimal position either remains constant or increases over iterations, as a position is selected as the global optimum only if it has a superior fitness value. Second, the minimum achievable rate for problem \eqref{2-9} is capped by an upper limit due to limited communication resources, which prevents indefinite increase. The PPSO method operates with a computational complexity of $\mathcal{O}\left(NKL_\mathrm{sum}\right)$, influenced by the total number of particles, $N$, the number of iterations, $K$, and the number of paths, $L_\mathrm{sum} = \sum\nolimits_{p \in \left\{ \mathrm{A,B} \right\}} {\sum\nolimits_{q \in \left\{ \mathrm{A,B} \right\}} {  {L_\mathrm{t}^{pq}L_\mathrm{r}^{pq} + L_\mathrm{t}^{pq}} } } $.

	In comparison, the convergence of the conventional APO method (which is similar to the block coordinate descent (BCD) method) is also guaranteed \cite{Ma_TWC_2023}. Its computational complexity is given by $\mathcal{O} \left( IGL_\mathrm{sum} \right)$, where $I$ is the number of iterations. We assume that each moving region is divided into $G$ grids to facilitate the exhaustive searches for the current MA. Generally, setting a larger value for $G$ is necessary to ensure the search accuracy of the APO method, e.g., each region of size $2 \lambda \times 2 \lambda$ is divided into multiple grids of size $\frac{\lambda}{100} \times \frac{\lambda}{100}$ \cite{Zhu_TWC_2023}, i.e., $G=200^2$, which results in a higher computational overhead compared to the PPSO method. Hence, the successive convex approximation (SCA) is employed in \cite{Ma_TWC_2023} to avoid exhaustive searches. Its computational complexity is given by $\mathcal{O} \left( I \hat I L_\mathrm{sum} \right)$, where $\hat I$ is the number of iterations for SCA.

	\section{Simulation Results}
	This section presents simulation results to illustrate the effectiveness of our proposed MA-enabled CCFD system. In the simulation, we adopt the geometry channel model as \cite{Zhu_TWC_2023,Ma_TWC_2023}, i.e., $L^{pq}_\mathrm{t} = L^{pq}_\mathrm{r} \triangleq L_{pq}$, in which the channel matrix ${\mathbf{\Sigma}_{pq}}$ is composed of $L_{pq}$ diagonal elements that follow the CSCG distribution $\mathcal{CN}\left( 0, \frac{\rho}{L_{pq}} \right)$ with $p = q$ for SI channel \cite{Guan_WCL_2022}, where $\rho =-90 \mathrm{dB}$ is the SI loss coefficient representing the path loss and the SI cancellation in both analog and digital domains, or follow the CSCG distribution $\mathcal{CN}\left( 0, \frac{\beta_{pq} d_{pq}^{-\alpha}}{L_{pq}} \right)$ with $p \ne q$ for SoI channel, where $\alpha=2.8$ is the path loss exponent, $\beta_{pq}=-30\mathrm{dB}$ is the path loss coefficient, and $d_{pq}=100\mathrm{m}$ is the propagation distance.  The elevation and azimuth AoDs and AoAs are assumed to be the independent and identically distributed (i.i.d.) random variables within the interval $\left[ -\frac{\pi}{2}, \frac{\pi}{2} \right]$. Unless otherwise stated, we set AWGN power $\sigma_\mathrm{q}^2=-80 \mathrm{dBm}$, transmit power $P_\mathrm{t}=20\mathrm{dBm}$, numbers of particles $N=200$ and iterations $K=100$, learning factors $c_1=c_2=1.4$, minimum/maximum inertia weight $\omega_{\min}/\omega_{\max}=0.4/0.9$, moving region size $D = \lambda$, and numbers of SI paths $L_{pq} = 5$ for $p = q$ and SoI paths $L_{pq} = 10$ for $p \ne q$.

	\begin{figure}
		\centering
		\subfloat[]{\label{fig3a}\includegraphics[width=0.5\columnwidth]{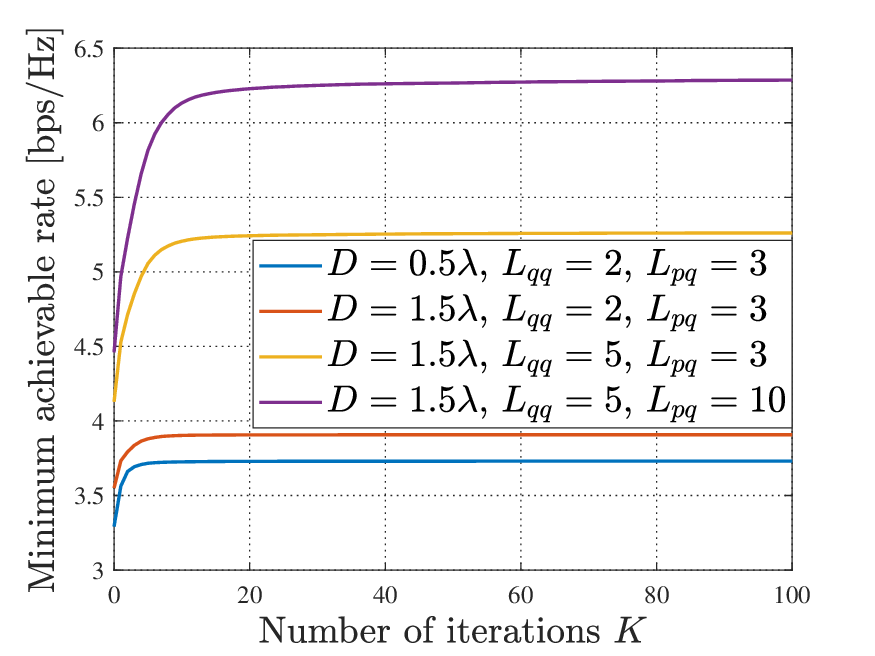}}
		\subfloat[]{\label{fig3b}\includegraphics[width=0.5\columnwidth]{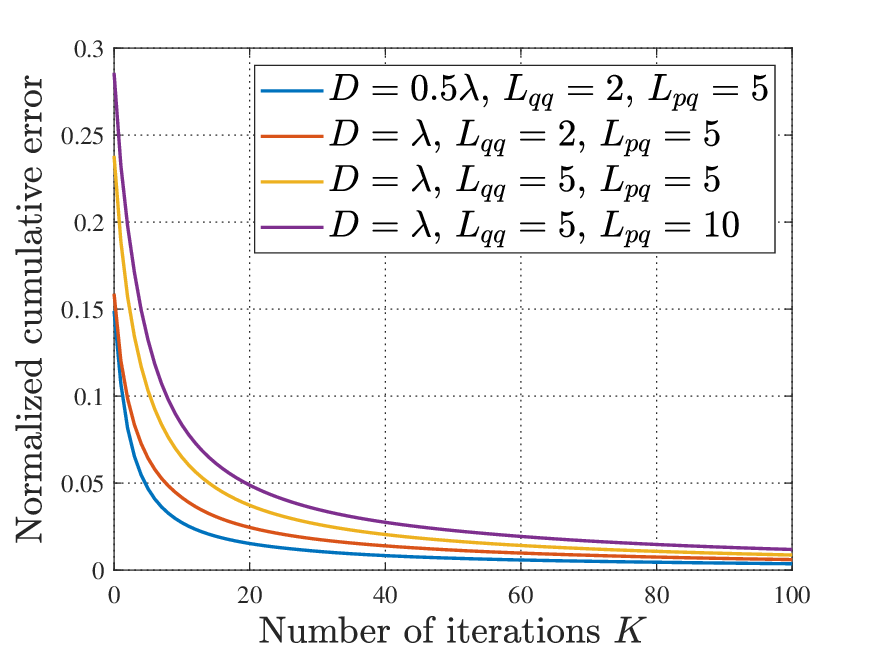}}\\
		\caption{Evaluations of the (a) convergence and (b) normalized cumulative error of the PPSO method.}
		\label{fig3}
	\end{figure}

	First, Fig.\;\ref{fig3} evaluates the convergence and normalized cumulative error of PPSO. As can be observed from Fig.\;\ref{fig3}\subref{fig3a}, the minimum achieve rates under various simulation settings increase with the number of iterations and the curves reach the steady states after 20 iterations. In Fig.\;\ref{fig3}\subref{fig3b}, the normalized cumulative error is defined as the sum of the differences between the optimized minimum achievable rate in each iteration ${F^{\left( k \right)}}\left( {{\mathbf{u}^*}} \right)$ and the desired value $F^*$, i.e., $\frac{1}{{\left( {K + 1} \right){F^*}}}\sum\nolimits_{k = 0}^K \left| {{F^*} - {F^{\left( k \right)}}\left( {{\mathbf{u}^*}} \right)}\right| $. We can observe that the normalized cumulative errors consistently decrease across different moving region sizes and numbers of paths, reaching approximately 0.01 after 100 iterations. This indicates that the PPSO effectively avoids undesired suboptimal solutions and progressively converges towards the desired value.

	\begin{figure}[!t]
		\centering
		\subfloat[SI channel gain]{\label{fig4a}\includegraphics[width=0.5\columnwidth]{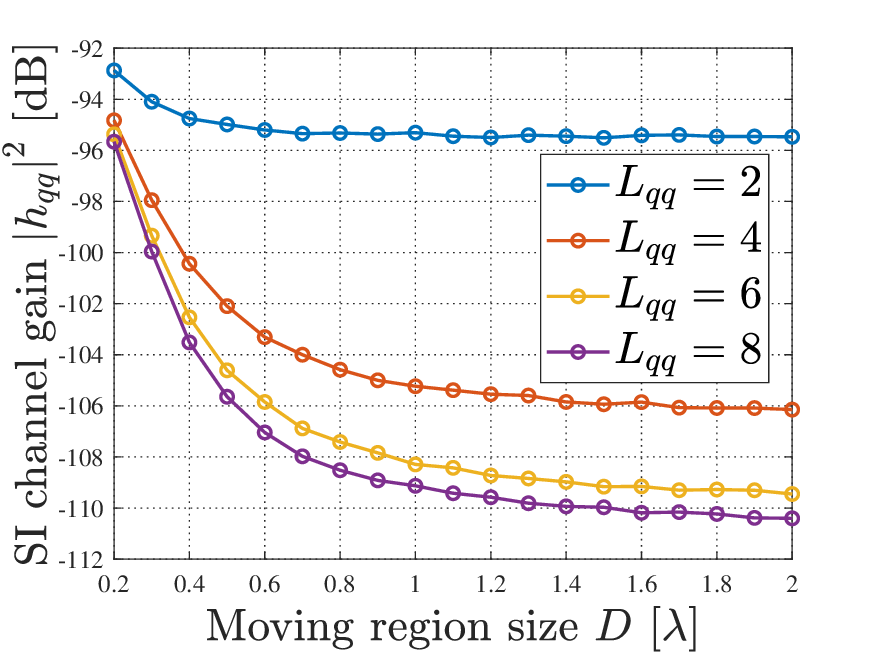}}
		\subfloat[SoI channel gain]{\label{fig4b}\includegraphics[width=0.5\columnwidth]{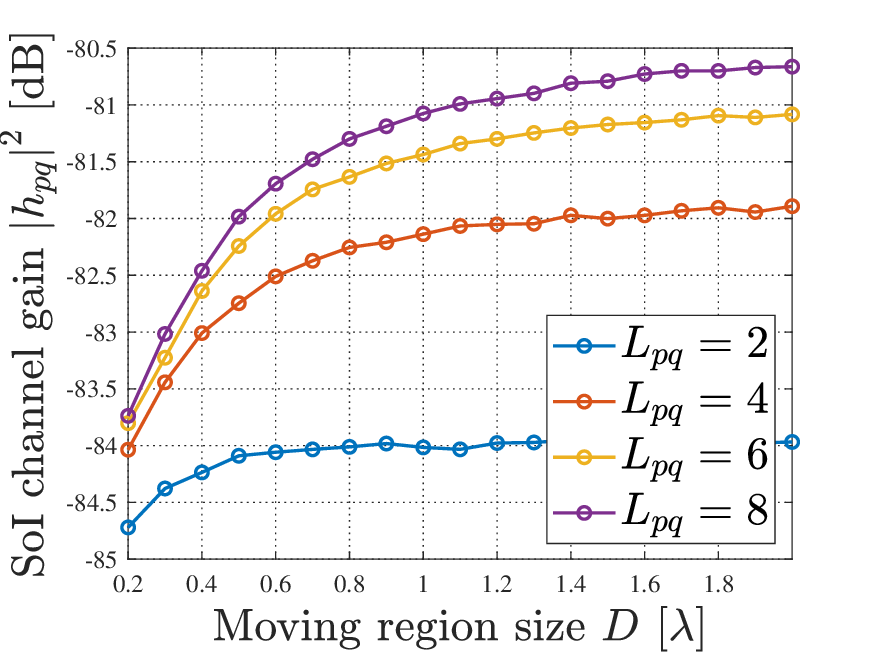}}\\
		\caption{Channel gains vs. moving region size $D$.}
		\label{fig4}
	\end{figure}

	Then, Fig.\;\ref{fig4} shows the SI and SoI channel gains versus the moving region size $D$ for different numbers of paths.  It can be seen that as the moving region size and the number of paths increase, the channel gains of SI and SoI increase and decrease, respectively.  This is because a larger moving region allows for more spatial DoFs to be explored, and more paths provide additional diversity gains.  Consequently, after optimization, MAs tend to settle in positions with lower SI channel gain and higher SoI channel gain to maximize SINRs.

	\begin{figure*}[!t]
		\centering
		\subfloat[]{\label{fig5a}\includegraphics[width=0.645\columnwidth]{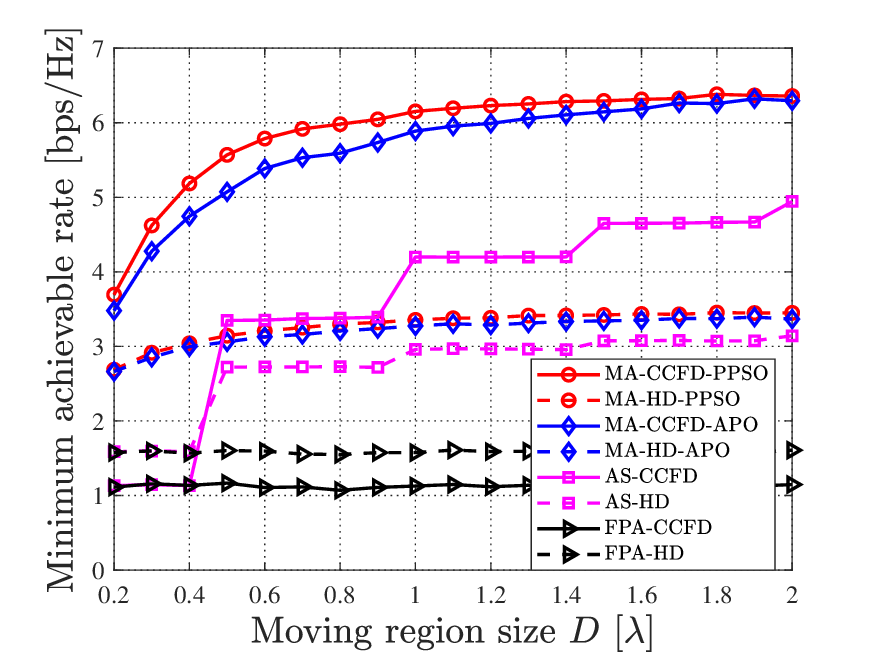}}
		\subfloat[]{\label{fig5b}\includegraphics[width=0.645\columnwidth]{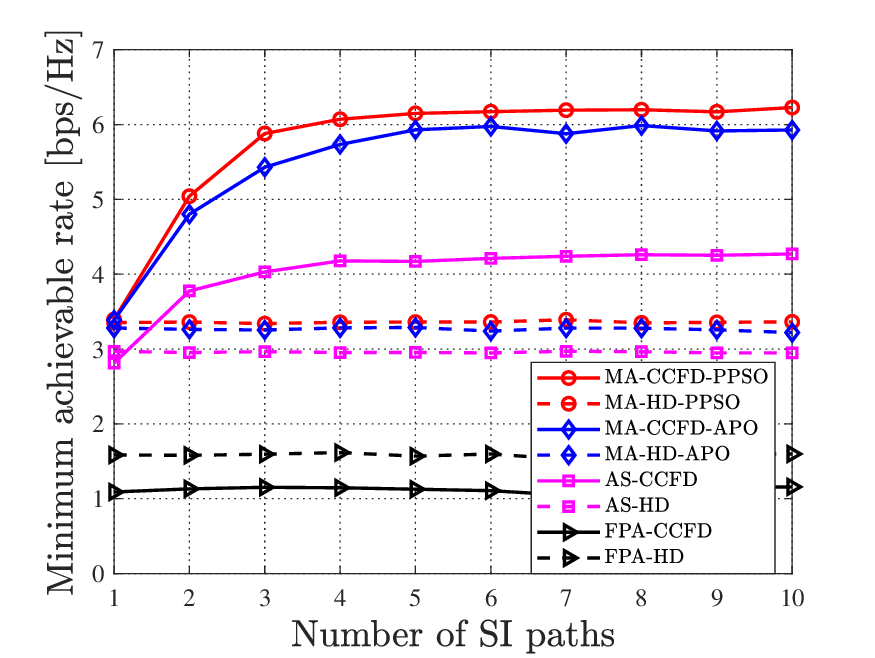}}
		\subfloat[]{\label{fig5c}\includegraphics[width=0.645\columnwidth]{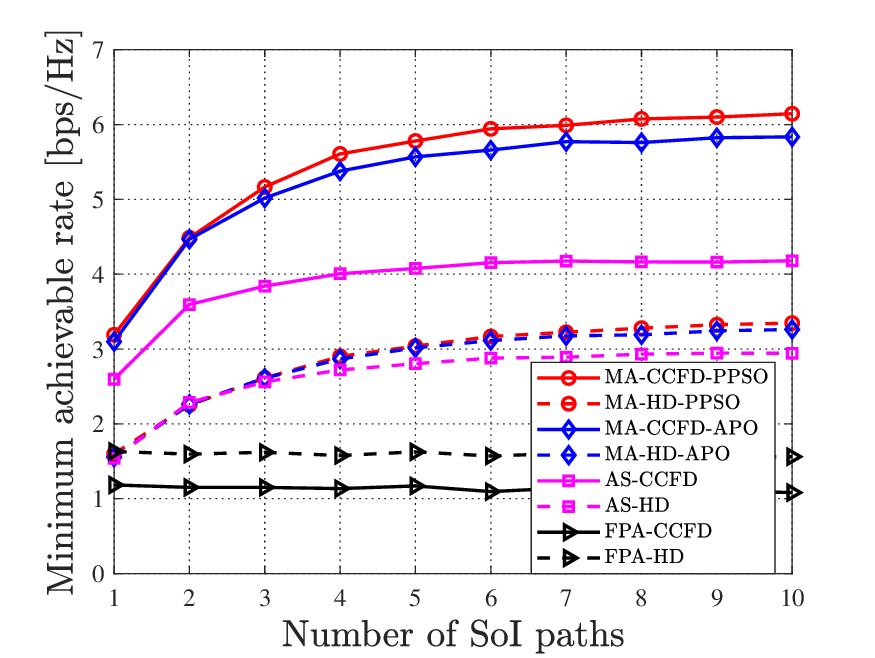}}\\
		\caption{Minimum achievable rates vs. (a) moving region size $D$, (b) number of SI paths, and (c) number of SoI paths.}
		\label{fig5}
	\end{figure*}

	Moreover, in Fig.\;\ref{fig5}, we compare the performance of various schemes, including MA, FPA, and AS, as well as CCFD and HD modes. In the AS-enabled scheme, the antenna arrays are deployed at intervals of $\frac{\lambda}{2}$ and the optimal antennas are selected through alternating optimization. In the HD mode, the achievable rate is penalized due to the half communication time, given by $R^\mathrm{HD}_q = \frac{1}{2}{\log_2}\left( 1 + \frac{{{{\left| {{h_{pq}}\left( {\mathbf{t}_p},{\mathbf{r}_q}\right) } \right|}^2}{P_\mathrm{t}}}}{{\sigma_q^2}} \right)$ with $~p\ne q$.  Also, we evaluate the performance of our proposed PPSO against the conventional APO method.  In the APO method, the moving region is partitioned into multiple grids of size $\frac{\lambda}{100} \times \frac{\lambda}{100}$, allowing for an alternative search of each MA's position while others remain fixed.   The antennas used in the aforementioned schemes cannot exceed the region of size $D \times D$. The proposed scheme is referred to as MA-CCFD-PPSO, signifying the MA-enabled CCFD scheme that utilizes the PPSO method.  Other schemes are labeled similarly.

	Fig.\;\ref{fig5}\subref{fig5a} investigates the minimum achievable rate concerning the moving region size $D$. It reveals that for MA-enabled schemes, the rates continuously rise as the moving region expands.  In contrast, in AS-enabled schemes, the increases in the rates are only noticeable when the number of deployed antennas is augmented and such an augmentation occurs if the size $D$ is an integer multiple of $\frac{\lambda}{2}$.  This is because the MA can more fully exploit the wireless channel spatial variation in a continuous region. Compared to FPA-enabled schemes, the MA and AS-enabled schemes demonstrate enhanced performance.  Moreover, the PPSO method surpasses the APO method, especially when $D$ is smaller than $\lambda$ in the CCFD mode.  However, for the HD mode, the improvement is less pronounced, indicating that the PPSO method can bring greater enhancements in scenarios of higher complexity.  Besides, it is interesting to observe that the AS-enabled HD scheme exhibits superior performance over the CCFD one when the size $D < \frac{\lambda}{2}$.  This superiority can be attributed to the deployment of a single antenna at the origin coordinate of the moving region, aligning its performance with that of the FPA-enabled schemes.

	Figs.\;\ref{fig5}\subref{fig5b} and \ref{fig5}\subref{fig5c} respectively depict the minimum achievable rates versus the numbers of SI and SoI paths.  In CCFD mode, it can be seen that the rates of both MA and AS-enabled schemes increase and gradually converge to the stable values with an increase in the number of SI or SoI paths. This is because more paths lead to higher multipath diversity gain but fully exploiting these gains requires a larger region. However, in HD mode, the rates hold steady when the number of SoI paths is constant, due to the absence of SI.  These results also imply that the PPSO method consistently outperforms the conventional APO method across different settings.  Furthermore, all the discussed schemes present superior performance over the FPA-enabled schemes.

	\section{Conclusion}
	In this paper, we proposed a point-to-point CCFD system enabled with MAs.  By dynamically repositioning antennas to simultaneously mitigate SI and enhance SoI, the MA-enabled CCFD system was capable of achieving an outstanding minimum achievable rate.  To effectively optimize the positions of MAs, the PPSO method was implemented, addressing the non-convex problem and preventing convergence to an undesired suboptimal solution. Simulation results confirmed the improvements offered by the PPSO method compared to the conventional APO method. Furthermore, these results highlighted the superiority of the MA-enabled CCFD scheme over the AS-enabled schemes and those deploying FPAs regardless of whether they operate in CCFD or HD modes.

\end{document}